\documentstyle[12pt]{article}
\def\Cstar{{\bf C}^\times}
\def\d{{\delta}}
\def\rb{{\displaystyle{\stackrel{\pi_i}
{\overrightarrow{\longrightarrow}}}}}

\def\~{\tilde}
\def\^{\hat}

\def\s #1{{\stackrel#1{\longrightarrow}}}

\def\H #1#2#3{{H^#1(#2,#3)}}

\def\sr #1 {{\stackrel{\partial_#1}{\longrightarrow}}}
\def\sl #1 {{\stackrel{\partial_#1}{\longleftarrow}}}
\def\e #1#2{{#1_1, \ldots , #1_#2}}

\def\ba{{\begin{array}}}
\def\ea{{\end{array}}}
\begin{document}
\begin{small}

\title{ Higher Bundle Gerbes and Cohomology Classes in Gauge Theories}

\author{ A.L.Carey \and  M.K.Murray \and B.L.Wang }

\maketitle
\centerline{ Department of Mathematics, University of Adelaide}
\centerline{Adelaide, SA 5005, AUSTRALIA}

\vskip1cm

\begin{abstract}

The notion of a higher bundle gerbe is introduced
to give a geometric realization of the
higher degree integral
cohomology of certain manifolds.
We consider examples using the infinite dimensional spaces
arising in gauge theories.

\end{abstract}

\section{Introduction}

This paper develops ideas hinted at in \cite{Mur} and \cite{CarMur}. In
order to make this account self contained we will review in Sections 2 and
3 below relevant aspects of these earlier papers. We are interested in the
general problem of realising higher degree cohomology classes of manifolds
geometrically. The work of Brylinski \cite{Bry} provides one approach to
this problem via a sheaf theoretic description of the category theorists
notion of a gerbe. In \cite{Mur} a simpler approach, which seems sufficient
for the applications we have in mind, was introduced by one of us. This
simpler notion of `bundle gerbe' enables us to realise classes in
$H^3(M,Z)$. The question of what to do with higher degree classes was posed
in \cite{Mur} and it was conjectured that a notion of bundle $n$-gerbe was
needed with 1-gerbes corresponding to the case described in \cite{CarMur}.
(Line bundles should be regarded as 0-gerbes in this setting.)

The main result of the present paper is to show that there is indeed a
notion of bundle $n$-gerbe (strictly speaking we discuss in detail the
general definition for 2-gerbes only). Our motivation comes from the
interesting examples \cite{Zum} of de Rham forms on the space of
connections obtained from Chern-Simons secondary characteristic classes
exploited in the physics literature to study the cohomology of gauge
groups. We do not completely resolve the connection between bundle gerbes
and the Chern-Simons classes here nevertheless we provide a convincing
application of the bundle $2$-gerbe notion using examples motivated by
gauge theories.

The main results of this paper can be summarised as follows. In Section 2
we extend the discussion of \cite{Mur} in two ways. First we give an
explicit proof of associativity of the product for bundle 1-gerbes in a
form which can easily be generalised to the 2-gerbe situation. We also
construct the tautological bundle 1-gerbe on manifolds which are not
2-connected. In Section 3 we develop  the definition of a bundle 2-gerbe.
This is related to the structures introduced in Freed \cite{Fre} in his
study of gauge theories. The discussion in Section 2 generalises in a
straightforward way to the case of bundle 2-gerbes. In Section 4 we review
\cite{CarMur} and then provide some examples, in the context of gauge field
theories, of the gerbe viewpoint. The most novel example is the case of
degree four de Rham cohomology of the space of connections modulo the gauge
action or equivalently via transgression of a 3-cocycle on the gauge group.
This is given a geometric realisation in this paper.

\noindent{\bf Acknowledgement}. We would like to thank Jouko Mickelsson
for discussions which led to the results
of the last two sections.

\section{Bundle 1-Gerbes}

In \cite{Mur}, bundle gerbes were developed to provide an
alternative  geometric realization
of three dimensional cohomology  to that given by Brylinski's
sheaf theoretic approach to gerbes \cite{Bry}.
In this paper we  generalise this definition to that of a bundle
$n$-gerbe which may be used to give  a geometric realisation of
cohomology in degree $n+2$.
First we review the  construction in \cite{CarMur}.    At various
points we refer to $H^p(M, Z)$ when what we really mean
is the image in the de Rham cohomology of the integral cohomology
of $M$.  We review first the theory of bundle gerbes.

\subsection{Review of bundle 1-gerbes.}

Consider a fibration  $\pi: Y \to M$. Denote by $Y_m$
the fibre of $Y$ over $m$, that is the set $\pi^{-1}(m)$. The fibre product
of $Y$ with itself, denoted  $Y^{[2]}=Y\times_\pi Y$, is a
new fibration whose fibre at $m$ is $(Y^{[2]})_m = Y_m \times Y_m$.
It is often useful to think of it as
the  subset of pairs $(y_1, y_2)$ in $Y\times Y$ such that
$\pi(y_1)=\pi(y_2)$.
A bundle gerbe is a principal $ C^\times $-bundle $P$ over $Y^{[2]}$ with
a composition map defined fiber by fiber smoothly as:
$$
\matrix{
P_{(x, y)} \times P_{(y, z)} &\longrightarrow P_{(x, z)}}
\eqno(2.1)
$$

This composition map (2.1) is a morphism of the bundle $P\otimes P$ over
the groupoid $Y^{[2]}\circ Y^{[2]} \longrightarrow Y^{[2]}$ which satisfies
associativity, where $Y^{[2]}\circ Y^{[2]} $
is the set of pairs $((y_1, y_2), (y_2, y_3))$. It is shown in \cite{Mur}
that a bundle gerbe also has an identity and an inverse. The
identity is a section of $P$ over the diagonal inside $Y^{[2]}$
and the inverse is a bundle map $P_{(x,y)} \to P_{(y,x)}$
denoted by $p \mapsto p^{-1}$  such that $(pz)^{-1} = p^{-1}z^{-1}$.

We denote a bundle gerbe on $M$ by the diagram
$$
\matrix{ P          &                   &                               \cr
         \downarrow &                   &                       \cr
         Y_1 ^{[2]} & \rb           & Y_1                       \cr
                                        &                       &
\downarrow    \cr
                                        &                       & M
\cr
}
\eqno (2.2)$$
where $\pi_i $ $(i=1, 2)$ denote the projections onto the various factors.

A bundle gerbe connection is a connection on $P\to Y^{[2]}$ which respects
the gerbe structure, this means that over the diagonal it is flat, that the
product
map on the bundle gerbe sends the connection to
itself and that the inverse map
sends the connection to its dual.  To understand how to extract the
Dixmier-Douady class of the bundle gerbe from this connection
we need to digress for a moment.  Given a  fibration
$Y \to M$ we can form repeated  fibre products  $Y^{[p]}$ whose
fibre at $m$ is $Y_m^p$. We call this the $p$th fibre product.
There are $p$ projections  $\pi_i \colon Y^{[p]} \to Y^{[p-1]}$,
each of which  just omits a factor.  Pulling back differential
$q$ forms with these and adding with  an alternating sign defines a
map $\delta\colon \Omega^q(Y^{[p]}) \to \Omega^q(Y^{[p-1]})$. This is in
fact a chain map (i.e $\delta^2 = 0$) and defines a complex:
$$
\Omega^q(M) \s{{\pi^*}}  \Omega^q(Y) \s{{\delta}} \Omega^q(Y^{[2]})
\s{{\delta} } \Omega^q(Y^{[3]}) \s{{\delta}} \cdots
\eqno(2.3)
$$
It was shown in \cite{Mur} that this complex is exact. The requirement
that the bundle gerbe connection be compatible with the product
implies that its curvature $F \in \Omega^2(Y^{[2]})$
satisfies $\delta(F) = 0$ and hence $F = \delta(f)$
 or $F = \pi_1^*f-\pi_2^*f$ for some two-form on $Y$. Then  we have
 $\delta(df) = d\delta(f)  = dF = 0$ so that $df = \pi^*(\omega)$
 where $\omega$ is a closed three form on $ M$.
 The three form
$\frac{1}{2\pi i}\omega $ defines an element in $H^3(M, Z)$
which is the Dixmier-Douady class of the gerbe \cite{Mur}.
If
$\frac{1}{2\pi i}\omega $ is cohomologous to zero in $H^3(M,  Z),$ then
$$
P \cong \pi_1^*(L)\otimes \pi_2^*(L^{-1})
\eqno(2.4)
$$
where $L$ is some $ C^\times $-bundle over $Y$. We call such a bundle gerbe,
a trivial bundle gerbe. For the details,  see \cite{Mur}.

A more cohomological treatment of this construction
can be obtained by considering the short sequence
$$
Y^{[2]}\rb Y^{[1]} \longrightarrow M
\eqno(2.5)
$$
which we can use to define a  short exact sequence of de Rham complexes:
$$
0\longrightarrow \Omega ^*(M)\s{{\pi^*}}\Omega^*(Y^{[1]}) \quad
\s{{\pi_1^*-\pi_2^*}}\quad
\Omega^*(Y^{[2]}) \bigcap Im (\pi_1^*-\pi_2^*)\longrightarrow 0
\eqno(2.6)
$$
This induces  a long exact sequence in cohomology:
$$
\cdots \longrightarrow H^q(M)\s{{\pi^*}} H^q(Y^{[1]})\s{{\pi_1^*-\pi_2^*}}
H_{\pi}^q(Y^{[2]})\s{{\triangle}}
H^{q+1}(M)\longrightarrow \cdots
\eqno(2.7)
$$
where $H_{\pi}^q(Y^{[2]})=\{ \omega \in H^q(Y^{[2]}) | \omega
=(\pi_1^*-\pi_2^*)f \quad \hbox{for} \quad f\in
\Omega^q(Y^{[1]})\}$. In fact, from the discussion above
we see that the Chern class of the bundle $Q \to Y^{[2]}_2$
is in $H^2_{\pi}( Y^{[2]}, Z) \subset H^2( Y^{[2]}, Z)$ and the
construction of the Dixmier-Douady class we described is
just the application of the map
$$
H^2_{\pi}( Y^{[2]}, Z)\s{{\triangle}} \H{{3}}{{M}}{{Z}}
\eqno(2.8)
$$
This is analogous to the Chern-map for line bundles.

When the manifold $M$ is 2-connected
there is an  explicit realization of a bundle gerbe associated with $\frac
{1}{2\pi i}
\omega \in H^3(M,  Z)$, called the tautological bundle gerbe \cite{Mur}.
We review that construction here in a slightly different form so that
we can prove explicitly the associativity not proved in \cite{Mur}.
Fix a basepoint $x_0$ in $M$ and
consider the based path fibration:
$$
Y={\cal P}_0 M =\{\rho :[0,1] \to M|\rho (0)=x_0\}
\eqno(2.9)
$$
$\pi : Y \to M $ , is given by $\pi (\rho )=\rho (1).$ Then $ Y^{[2]}$
is the space of pairs of  smooth paths starting from $x_0$ and ending with
the same end point.
 We can construct a $\Cstar$ bundle $Q$  over $ Y^{[2]}$
by defining the fibre at $(\rho_0, \rho_1)$ to be the space whose
elements are equivalence classes of pairs $[\mu, z]$ where $\mu \colon [0,1]
\times [0,1] \to X$
is a piece-wise smooth homotopy, with endpoints fixed, from $\rho_0$ to
$\rho_1$ and $z$ is a
non-zero complex number. Recall that a
homotopy with endpoints fixed satisfies,
$\mu(s, 0) = \rho_0(0) = \rho_1(0) = x_0$,
$\mu(s, 1) = \rho_0(1) = \rho_1(1)$
for all $s$ and $\mu(0, t) = \rho_0(t) $ and $\mu(1, t) = \rho_1(t)$
for all $t$.  We say two pairs  $(\mu, z)$ and $(\mu', z')$ are
equivalent if for any homotopy
$$
F \colon [0,1] \times [0,1] \times [0,1] \to M
$$
between $\mu$ and $\mu'$
we have $\exp(\int F^*(\omega)) z = z'$, where the integral is over
$[0,1] \times [0,1] \times [0,1] $. The condition on the
homotopy $F$
is that $F(0, s, t) = \mu(s, t) $, $F(1, s, t) = \mu'(s, t) $ and for
each $r$
we have that $F(r, \ , \ )$ is a homotopy with endpoints fixed between
$\rho_0$ and $\rho_1$.

We want to now construct the product
$$
Q_{(\rho_1, \rho_2)} \otimes Q_{(\rho_2, \rho_3)} \to Q_{(\rho_1, \rho_3)}.
$$
If $\alpha$ is a homotopy from $\rho_1$ to $\rho_2$ and
$\beta$ is a homotopy from $\rho_2$ to $\rho_3$ then we can
construct a homotopy $\alpha\circ\beta$ from $\rho_1$ to
$\rho_2$ in the usual way by letting $\alpha\circ\beta(s, t) $
equal $\alpha(2s, t)$ for $s$ between $0$ and $1/2$ and all $t$
and letting $\alpha\circ\beta(s, t) $ equal $\beta(2s-1, t)$ for
$s$ between $1/2$ and $1$ and all $t$.
We need to check that this map is well-defined, that is
it respects the equivalence relation defining $Q$.
Consider then $\alpha'$ and $\beta'$ with $F$ a homotopy
from $\alpha$ to $\alpha'$ and $G$ a homotopy
from $\beta$ to $\beta'$. We can construct a homotopy $F*G$
from $\alpha\circ\beta$ to $\alpha'\circ\beta'$ by letting
$$
(F\circ G)(r,\ ,\ ) = F(r,\ ,\ )\circ G(r,\ ,\ )
$$
for all $r$. Because the linear change defining $\circ$ is
a diffeomorphism we deduce that
$$
\int (F\circ G)^*(\omega) = \int (F)^*(\omega)  + \int (G)^*(\omega)
$$
and it follows that if $[\alpha, z] = [\alpha', z']$
and $[\beta, z] = [\beta', w']$ then
$[\alpha*\beta, zw] = [\alpha'*\beta', z'w']$.
The product on $Q$ is then defined by
$$
[\alpha, z] \otimes [\beta, z]  \mapsto [\alpha\circ \beta, zw].
$$

We want to prove that this product is associative.  Let $\rho_4$
be another path and let $\gamma$ be a homotopy between $\rho_3$ and
$\rho_4$. It suffices to prove that $(\alpha \circ\beta ) \circ \gamma$
and $\alpha\circ(\beta\circ\gamma)$ are homotopic to each
other by a homotopy $F$ satisfying $F^*(\omega) = 0$.
To construct the homotopy let $r$ be a point in $[0,1]$ and
consider the division of  $[0, 1]$ into three intervals
$$
[0, (1+r)(1/4)], \quad [(1+r)(1/4), (1+r)(1/2)], \quad  [(1+r)(1/2). 1]
$$
and let $m_r(\alpha, \beta, \gamma)(\ ,t)$ be the homotopy obtained
 by applying $\alpha(\ ,t)$ $\beta(\ ,t)$
and $\gamma(\ ,t)$  at an appropriately
scaled speed to each of these intervals respectively.
So, in particular, $m_0(\alpha, \beta, \gamma) =
(\alpha\circ\beta)\circ\gamma$
and $m_1(\alpha, \beta, \gamma) = \alpha\circ(\beta\circ\gamma)$. Then
define
$$
F(r, s, t ) = m_r(\alpha, \beta, \gamma)(s  ,t).
$$
Notice that the image of $F$ is, at best, two-dimensional
so that the pull-back of the three form $\omega$ is
zero, as required. In fact it is not difficult
to define the lines in $[0,1 ] \times [0,1] \times [0,1]$
along which $F$ is constant.  It follows
that the product is associative as is required to
define a bundle gerbe.

The well-known example of a bundle gerbe is given by
the central extension of the loop
group \cite{PreSeg}, which is a realization of a degree 3
cohomology element, namely the generator of $\H{{3}}{{G}}{{Z}}$:
$$
(\xi_1, \xi_2, \xi_3)\longmapsto \frac {1}{8\pi^2}<\xi_1, [\xi_2, \xi_3]>
\eqno(2.10)
$$

\subsection{The tautological construction: non-connected case.}

When
 the manifold $M$ is not 2-connected, we cannot apply the above procedure to
construct the tautological bundle gerbe. But if we assume that
$\pi_2 (M) $ has no non-trivial $C^\times$-extensions, we can still
realize  the tautological bundle gerbe construction
 in a slightly different way.

Now it is not difficult to see that we may
identify the space $ Y^{[2]}$ of $\cal P_0 (M)$ over $M$ as the loop
space $S^1(M)$ of piecewise smooth loops in $M$.
 We may assume that $M$ is simply-connected (for if it is
not then we simply perform the construction below over each
connected component of $S^1(M)$).
 Through the evaluation map, we can pull back the three
form $\omega $ on $M$ to $S^1(M) \times [0, 1]$ and integrate out $[0, 1]$
to obtain an integral two form $\Theta$ on $S^1(M)$. Note that $\pi_1( S^1(M))
=\pi_2(M)$. Denote  the universal covering space  as $\widetilde S^1(M)$.
Pull back $\Theta $ to $\widetilde S^1(M)$, and denote
the pulled back form by $\widetilde \Theta$, which is
$\pi_2(M)$-invariant. Since $\widetilde S^1(M)$ is simply connected, we
can construct a $C^\times$-bundle $\widetilde Q$
 over $\widetilde S^1(M)$ by the
tautological method. Now it is fairly standard in the theory of line bundles
(see however \cite{CarPal} for a discussion) that the fundamental group of
$\widetilde S^1(M)$ (here $\pi_2 (M)$ ) has a $C^\times$-extension
$\widetilde \pi_2 (M)$, which acts on $\widetilde Q$. By assumption,
 $\pi_2 (M)$ has only trivial $C^\times$-extensions, therefore we can
view $\pi_2 (M)$ as a subgroup in $\widetilde \pi_2 (M)$, quotient
$\widetilde Q$ by this subgroup, we get a $C^\times$-bundle $Q$ over $S^1(M)$.

To verify that the quotient bundle is a bundle gerbe, we use the expression
for the lifted action of $\pi_2 (M)$ \cite{CarPal}.
 Recall that the universal cover
of  $S^1(M)$, $\widetilde S^1(M)$, consists of the based homotopic disks
 in $M$, that
is, over the point $\gamma \in  S^1(M)$, the fiber is the homotopic disk
bounded by $\gamma $. There is a natural
 commutative diagram which gives the groupoid
structure in $\widetilde S^1(M)$,
$$\matrix{
\widetilde S^1(M) & \circ & \widetilde S^1(M) & \longrightarrow
&\widetilde S^1(M) \cr
&  \downarrow  &    &    &\downarrow \cr
S^1(M) &  \circ &  S^1(M) &\longrightarrow&   S^1(M) \cr
}
$$
The line bundle $\widetilde Q$ is given by
triples $([D_1], p_{[D_1]}, z_1) $ modulo the equivalence relation given by
the two form $\widetilde \Theta$, where $ p_{[D_1]}$ is the path from the
base point $[D_0]$ in $\widetilde S^1(M) $ to $[D_1]$. For $\psi \in
\pi_1( S^1(M)) =\pi_2(M)$, the lifted action (under the assumption that
$\pi_2 (M) $ has no non-trivial $C^\times$-extension) $\hat \psi $
is calculated in \cite{CarPal} as follows,
$$
\hat \psi . ([D_1], p_{[D_1]}, z_1)
= (\psi([D_1]),p_\psi*\psi( p_{[D_1]}), z_1)
$$
where $p_\psi$ is the fixed path (only depending on $\psi$) from $[D_0]$
to $\psi([D_0])$ which is defined by the splitting map for the exact
sequence:
$$
0 \rightarrow C^\times \rightarrow \widetilde \pi_2 (M) \rightarrow\pi_2 (M)
\rightarrow 0.
$$

The product structure on $\widetilde Q$ is given by
$$
([D_1], p_{[D_1]}, z_1) \circ ([D_2], p_{[D_2]}, z_2)
=([D_1\circ D_2],  p_{[D_1]}\circ p_{[D_2]}, z_1z_2)
.$$
We only need to prove that the above product structure on $\widetilde Q$
is a  $\pi_2 (M)$-homomorphism. This is just the following
identity for the groupoid structure on $ \widetilde S^1(M)$:
$$
p_\psi*\psi( p_{[D_1]}\circ p_{[D_2]}) =
p_\psi*\psi( p_{[D_1]})\circ p_\psi*\psi( p_{[D_2]})
.$$
Therefore the gerbe structure on $ \widetilde S^1(M)$
descends to a gerbe structure on $S^1(M)$.

\section{Bundle 2-gerbes}

To generalise these ideas to higher gerbes we note
that   $Y^{[2]}$ is just the space $S^1(M)$
of based loops in $M$. The two projection maps project a loop
to the two paths corresponding to restricting to the
upper and lower \lq hemispheres'  of $S^1$.  The usual
evaluation map
$$
S^1(M) \times [0,1] \to M
\eqno(3.1)
$$
allows us to pull back $\omega$ and integrate out the
$[0,1]$ to obtain a two-form on $S^1(M)$. This
is the  curvature considered above.

Hence (2.2) becomes the diagram:
$$
\matrix{
Q                       &                       &                \cr
\downarrow      &                       &                         \cr
S^1(M)      &\rb & D^1(M)         \cr
                        &                       &\downarrow
\cr
                        &                       &M .         \cr
}
\eqno(3.2)
$$

An obvious generalisation of this which works for
3-connected manifolds $M$ is to consider the
diagram:
$$
\matrix{
Q                       &                       &                       &
&                       \cr
\downarrow      &                       &                       &
&                       \cr
S^2(M)      &\rb & D^2(M)           &                       &
\cr
                        &                       &\downarrow     &
&                       \cr
                        &                       &S^1(M)      &\rb
&D^1(M)            \cr
                        &                       &                       &
&\downarrow     \cr
                        &                       &                       &
&M.                  \cr
}
\eqno(3.3)
$$
Here
$S^n(M)$ is the space of based  maps from  $n$ spheres to  $M$
 with the base point on the equator and $D^n(M)$ the space of based
 maps from the $n$-dimensional ball to $M$ with the
 base point on the boundary.
  By restricting to the upper and lower hemispheres
of $S^n$ we obtain a pair of projections
$$
S^n(M) \rb D^n(M).
\eqno(3.4)
$$
Restricting to the boundary  of the $n$ disk also defines a map
$$
\pi \colon D^n(M) \to S^{n-1}(M).
\eqno(3.5)
$$
This means that $D^n(M)$ is a fiber bundle over $S^{n-1}(M)$ whose fiber
$D^n_f (M)$ at $f: S^{n-1} \rightarrow M$ is all
the extensions of $f$ to the $n+1$-dimension ball
 and we have that
 $$
 S^{n}(M) = D^n(M) \times_\pi D^n(M).
 \eqno(3.6)
 $$
The induced projection $S^{n}(M) \to S^{n-1}(M)$ is
that induced by restriction to the equator.  The line bundle $Q$
is defined in a manner analogous to the bundle 1-gerbe case or
equivalently we use the evaluation map
$$
S^2(M) \times S^2 \to M.
\eqno(3.7)
$$
Pullback with this and integrate over the $S^2$ to define a two-form $F$ on
$S^2(M)$.
This is closed and integral and hence defines a line bundle which is $Q$.
We call (3.3) the tautological bundle 2-gerbe.

Now, from the viewpoint of de Rahm cohomology the construction
above works as follows.
First we  construct a sequence of forms by starting with
$\Theta\in H^4(M,Z)$, then set
$$
\begin{array}{ll}
& F= \int_{S^2} ev^*(\Theta )\in H^2(S ^2(M), Z),\\[2mm]
& f_1 = \int_{D^2} ev^*(\Theta )\in \Omega^2( D^2 (M)),\\[2mm]
& \omega =\int_{S^1} ev^*(\Theta )\in H^3(S ^1 (M), Z),\\[2mm]
& f_2 = \int_{D^1} ev^*(\Theta )\in \Omega ^3( D^1 (M)).
\end{array}
$$

It is easy to show that the above forms satisfy:
$$
\begin{array}{ll}
& F= (\pi^*_1 - \pi^*_2 )f_1,\\[2mm]
& df_1 = \pi^*(\omega), \\[2mm]
& \omega = (\pi^*_1 -\pi^*_2 )f_2,\\[2mm]
& df_2 = \pi ^*(\Theta ).
\end{array}
\eqno(3.8)
$$

Recalling the definition of $Q$ above, $Q= D^3(M) \times C^\times /\sim$ where
the equivalence relation is given by
$$
(B_1, z_1 )\sim (B_2, z_2)
$$
if and only if $\partial B_1 = \partial B_2  \in S ^2(M)$
and
$z_1 = z_2 \exp(\int _{D^4} \Theta )$, with $D^4$ being the 4-dimensional disk
bounded by  the 3-sphere
formed by gluing $B_1$ and $B_2$ along the common boundary.
 One may easily check
that the definition is independent of the choice of $D^4$.
It is now straightforward,  using the methods of Section 2,
to define a product on $Q$ and show that  $Q$ is a bundle gerbe over
$S^1(M)$ whose Dixmier-Douady class is $\omega =
\int_{S^1} ev^*(\Theta )$ in $H^3(S ^1 (M), Z)$ (from the first two
identities in (3.8)).   Indeed it suffices really to note that
$D^2(M) $ is the space of based paths in $S^1(M)$ and that the
rest of the construction is just the tautological bundle
gerbe for the case of the three form  $\omega =\int_{S^1} ev^*(\Theta )$.

In the present situation as well as the bundle gerbe structure
there is additional structure in the form of
a multiplication on the fibration $D^2 (M) \rightarrow
S ^1(M)$ which can be lifted to $Q$.  Specifically, for two
loops $(\gamma_1 ,\gamma_2)$ and $(\gamma_2, \gamma_3) $ in $ S^1(M)$,
one can define
$$
(\gamma_1 ,\gamma_2) * (\gamma_2, \gamma_3)  = (\gamma_1, \gamma_3).
$$

This product can be lifted to the total space of
the fibration $D^2 (M) \rightarrow
S ^1(M)$ as follows.  If we think of
two bounding disks in $D^2(M)$ as homotopies
$\mu$ and $\rho$ from $\gamma_1$ to $\gamma_2$ and
$\gamma_2$ to $\gamma_3$ then we can compose them as
in Section 1 to get a homotopy $\mu*\rho$ from
$\gamma_1$ to $\gamma_3$.   Hence we have a  product:
$$
m:  D^2(M) _{\gamma_1}  \times D^2(M)_{\gamma_2}  \longrightarrow
D^2(M)_{\gamma_1 \circ \gamma_2 }.
$$
Note that this product is not associative.
The product $m$ also defines a product
$$
m:  S^2(M) _{\gamma_1}  \times S^2(M)_{\gamma_2}  \longrightarrow
S^2(M)_{\gamma_1 \circ \gamma_2 }.
$$
 This product can be lifted to a product on $Q$ as follows.
Let $[B_1, z_1] \in Q_{(\mu_1, \rho_1)} $ and $[B_2, z_2] \in Q_{(\mu_2
\rho_2)}$.
Then we can think of $B_i$ as a homotopy from $\mu_i$ to $\rho_i$
and we can define a homotopy from $\mu_1*\rho_1$ to $\mu_2*\rho_2$
by $(B_1*B_2)(s,\ ) = B_1(s,\ )*B(s,\ )$. The product is then
$$
[B_1, z_1 ] *  [B_2, z_2] = [B_1 * B_2, z_1z_2].
$$
It is straightforward (but very tedious)
using  the methods of Section 2 to check
that this product is well-defined and associative.  In a similar fashion
we can show that  this product covers the product $m$
above. So, we have lifted the product on $S^1(M)$ to
a product on the bundle gerbe over $S^1(M)$. It is therefore
natural to introduce the following:

\medskip
\noindent{\bf Definition 3.2}
A  bundle $2$-gerbe is a   diagram of spaces of
the form
$$
\matrix{
Q                       &                       &                       &
&                       \cr
\downarrow      &                       &                       &
&                       \cr
Y^{[2]}_2      &\rb & Y_2           &                       &
\cr
                        &                       &\downarrow     &
&                       \cr
                        &                       &Y_1^{[2]}      &\rb
&Y_1           \cr
                        &                       &                       &
&\downarrow     \cr
                        &                       &                       &
&M                   \cr
}
\eqno(2.23)
$$
where
$$
\matrix{ Q          &                   &                               \cr
         \downarrow &                   &                       \cr
         Y_2 ^{[2]} & \rb           & Y_2                      \cr
                                        &                       &
\downarrow    \cr
                                        &                       & Y^{[2]}_1
\cr
}
\eqno(3.9)
$$
is required to be a bundle 1-gerbe and the natural product on
$Y^{[2]}_1$ is covered by the product on $Q$.
That is, there is a fibration composition map
over the groupoid
$Y_1^{[2]} \circ  Y_1^{[2]} \longrightarrow Y_1^{[2]}$
defined fiber by fiber as
$$
m: \quad (Y_2) _{(y_1, y_2)} \times (Y_2) _{(y_2, y_3)} \longrightarrow
(Y_2)_{(y_1, y_3)}
\eqno(3.10)
$$
where $(y_1, y_2), (y_2, y_3) \in Y_1^{[2]}$ and this map $m$
is compatible with the multiplication in $Q$.

It is clear now that we could extend the tower of spaces
in (3.9) and define bundle $n$-gerbes however we leave this refinement
to the reader (note this is not a trivial extension as
one needs to keep track of the product at each level).
 Finally we note that we may handle the non-connected
tautological bundle 2-gerbe construction in a fashion
similar to that for the bundle 1-gerbe. The space $D^2(M)$ is defined in the
 obvious fashion over each connected component of $S^1(M)$.
One defines a gerbe structure using the WZW construction on the
simply connected covering space of (each component) of $S^2(M)$
and under the assumption that the fundamental group of
$S^2(M)$ has no non-trivial $C^\times$-extensions one may factor out
by an action of this fundamental group to obtain the
tautological bundle 2-gerbe.

\section{ Transgression}

In \cite{CarMur} two of us introduced a twist on the usual transgression map
arising from  the contractibilty of the space of connections on a principal
bundle.
In our approach we use transgression to define a map
 from the $p^{\mbox{th}}$-cohomology of connection space
modulo gauge
transformations to the $(p-1)^{\mbox{th}}$-cohomology of the
 Lie algebra of the gauge group with coefficients in functions on the
space of connections. This unifies the two views of anomalies as
manifestations of the non-trivial topology of the space of connections modulo
gauge transformations on the one hand and
group cohomology of the gauge group on the other.

Let $(P, G, M)$ be the $G$-principal bundle over a compact Riemannian
manifold with a compact structure group $G$, let $\cal A$ denote the affine
space of connections, modelled on  $Ad P$-valued 1-forms on $M$, $\Omega
^1_M(Ad P)$, where the $Ad P$ is the associated bundle by the adjoint
representation of $G$ on its Lie algebra $L(G)$. Denote by $\cal G$ the
gauge group, viewed as the automorphisms of $P$, respecting the fibres and
covering the identity map of $M$, $L(\cal G )$ its Lie algebra which can be
identified with $\Omega _M^0(Ad P)$ the $Ad P$-valued functions on $M$. By
suitable basepointing, $\cal G$ acts on $\cal A$ freely.

In \cite{CarMur}, we introduced the transgression map:
$$
\H{p}{{{\cal A}/{\cal G}}}{R}\longrightarrow \H{{p-1}}{{L({\cal
G})}}{{Map({\cal A}, R)}}
\eqno(4.1)
$$
where $\H{p}{{{\cal A}/{\cal G}}}{R}$ is the $p$-th de Rham cohomology
group of the moduli space ${\cal A}/{\cal G}$,  and the notation
$$\H{{p-1}}{{L({\cal G})}}{{Map({\cal A}, R)}}$$ means the
$(p-1)^{\mbox{th}}$-cohomology group of the  Lie algebra $L({\cal G})$ with
values in $Map({\cal A}, R)$. This transgression map is defined as follows.
If $\omega \in \Omega ^p({\cal A}/{\cal G})$ with $d\omega =0$, then the
pull back $\pi^*\omega$ in $\Omega^p({\cal A}) $ is an exact form on $\cal
A$ due to the fact that $\cal A$ is an affine space. So $\pi ^*\omega = \d
\mu$ where $\d$ is the exterior differential operator on $\cal A$ and $\mu
\in \Omega ^{p-1}({\cal A})$.
Consider ${\cal A}=\bigcup_{\cal A}{\cal G}.A$, then the vector field
induced by the
infinitesimal gauge  transformation $\epsilon $ is $-d_A \epsilon $ and the
$p-1$ cocycle corresponding to
$\omega $ is given by
$$
(\e{{\epsilon}}{{p-1}})\longmapsto (-1)^{p-1}\mu (\e{{d_A\epsilon}}{{p-1}}).
\eqno(4.2)
$$
The Lie algebra coboundary operator corresponds to the exterior
differential operator on the de Rham
complex and so $(4.2)$ induces the transgression map $(4.1)$.

\subsection{The Atiyah-Singer Construction of closed forms on $\cal A/\cal G$}

 There is a universal bundle $\cal L$ over $M \times (\cal A /\cal G)$
\cite{A-S, D-K} described as follows.
At $(p, A) \in M \times \cal A $, the connection $A$ gives a
decomposition of the tangent space $T_p(P) = T_{[p]}(M) \oplus Lie(G)$
which gives a  $G$-Riemannian metric on $P$, this metric and the correspoding
Hodge $*$ operator endows $M \times \cal A $
with a Riemannian metric which is
invariant under $\cal G \times G$.
Now construct $\cal L = (M \times \cal A)/\cal G$
then the universal bundle is $\cal L $ with base
space $\cal L /G$ with its
natural connection $\theta $. The  curvature $\cal F$
of $\theta$ has several  components
according to the degree in the $M$ and $\cal A /\cal G$ directions.

Recall that the tangent space of $\cal A/\cal G$ at $[A]$
consists of those
$Lie (G)$-valued 1-forms on $P$ which lie in the kernel of $D_A^*$.
Next we note that
$$
\cal F = \cal F^{2,0} + \cal F^{1,1} +\cal F^{0,2}
$$
where

(1)$ \cal F^{2,0} _{(x, A)} = F_A$,

(2) $\cal F^{1,1}_{(x, A)} (X, \xi) = \xi (X)$
for $X \in T_x(M), \xi \in T_{[A]}(\cal A/\cal G)$,

(3) $\cal F^{0,2}_{(x, A)} (\xi, \eta ) = -(D_A^* D_A)^{-1}(*
( \xi \wedge *\eta ))$
where $\xi, \eta  \in  T_{[A]}(\cal A/\cal G)$.

For dim $M = 3$, we can construct a closed  3-form
on $\cal A/\cal G$ by considering
$$\int _M tr (\cal F)^3. \eqno (4.3) $$
The only contribution is from the terms
$$
2 \mbox{str}(\cal F^{2,0} \wedge \cal F^{1,1}\wedge \cal F^{0,2}) +
tr ( \cal F^{1,1})^3.
$$

For the case dim $M = 4$ we again construct a closed 4-form
 on $\cal A/\cal G$ by integration
$$\int _M tr ( \cal F)^4 \eqno (4.4)$$
and the only contributing terms are
$$
\mbox{ str} ( F^{0,2} \wedge F^{0,2} \wedge\cal F^{0,2}\wedge \cal F^{0,2})
+ tr ( \cal F^{1,1})^4 +\mbox{str}
 ( F^{2,0} \wedge \cal F^{1,1}\wedge \cal F^{1,1}
\wedge \cal F^{0,2} ). $$
(Here str means the symmetric trace.
Suitably normalised (4.3) and (4.4)  define
forms $\Theta_n, n =3,4$ which determine integral
cohomology classes on $\cal A/\cal G$.
Then, using the viewpoint of higher bundle gerbes, $\Theta _n$ defines
the tautological bundle $n-2$-gerbe on ${\cal A}/{\cal G}$.
We discuss the $n=3$ cases in more detail in the next subsection
and the $n=4$ case in section 5.

\subsection{ A 1-Gerbe on ${\cal A}/{\cal G}$  and the
Faddeev-Mickelsson Cocycle}

In this subsection, we give some details of the
 tautological bundle 1-gerbe
derived from the form (4.3) and give various geometric
interpretations including
relating it to the
 Faddeev-Mickelsson cocycle which gives rise to
an extension of the gauge group.

Suppose $M$ is a 3-dimensional compact closed manifold
with $\pi_2(M)$ satisfying the constraint of subsection 2.2
(for the sake of concreteness one may take $M$ to be $S^3$).
 The degree 3 form $\Theta _3$ of equation (4.3),
which gives a class in $ H^3({\cal A}/{\cal G})$,  pulls back to $\cal A$.
We then transgress to obtain
the corresponding 2-cocycle on $L({\cal G})$.
In \cite{cmm} we showed that the resulting cocycle
was cohomologous to the Faddeev-Mickelsson cocycle
\cite{Seg}, \cite{Fad} \cite {Mic} here denoted $\mu_2$:
$$
\mu_2 (\epsilon_1, \epsilon_2) =-\frac{i}{2\pi^3}\int_M sTr (A\wedge
d_A\epsilon_1\wedge d_A\epsilon_2)
\eqno(5.1)$$
where $\epsilon_i \in L({\cal G})$.

Now by the construction of subsection 2.2
$ \Theta _3$ defines the tautological
bundle 1-gerbe on ${\cal A}/{\cal G}$:
$$
Q_1 \longrightarrow Y^{[2]}\rb Y^{[1]}={\cal P}({\cal A}/{\cal
G})\longrightarrow {\cal A}/{\cal G}
\eqno(5.2)
$$
Here $
Q_1 \longrightarrow Y^{[2]}$ is a line bundle over the
 the loop space
of ${\cal A}/{\cal G}$, $S^1({\cal A}/{\cal G})$,  with first Chern
class given by
$$
2\pi i \int_{S^1}\! ev^*(\Theta_3 )= \omega _2
\eqno(4.5)
$$
where $ev: S^1({\cal A/\cal G}) \times S^1 \longrightarrow {\cal A}/{\cal G}$
is the evaluation map.

We may also pullback $\omega_2$ to $\cal A$ (we denote the pullback form
by the same symbol).
 It then defines a
trivial bundle 1-gerbe on $\cal A$
which is a line bundle on $Y^{[2]}=S^1(\cal A )$ given explicitly as follows.
Take triples consisting of a loop
 $A \in S^1(\cal A) $, a path $\gamma$ joining a
fixed base point to $A$ and
$z\in C$ (note that we need to fix a base point in
 each connected component). Consider equivalence classes:
$$
[(A, \gamma , z)]
\eqno(4.6)$$
under the equivalence relation given by
$$
(A_1, \gamma _1, z_1) \sim (A_2,\gamma_2, z_2)
\eqno(4.7)
$$
if and only if $A_1 \equiv A_2$, and $z_1=z_2 \exp (\int_\sigma \omega_2)$,
where $\sigma $ is a surface in
$\cal A$ with boundary $\gamma _1 * \gamma _2^{-1}$.
An easy calculation reveals that the
 gauge group $\cal G$ acts on this
line bundle by
$$
g.[( A, \gamma , u)]= [(g.A, g.\gamma , u \exp
(\int _\gamma \alpha )\!)].
\eqno(4.8)
$$
Since $S^1(\cal A)$  is also affine, we have
$$
\omega _2(A)=\d \rho (A)
\eqno(4.9)
$$
where $\rho$ is a 1-form on $S^1(\cal A)$. Moreover
$ \rho  \vert_{S^1(\cal G) .A}$
defines the same cohomology class as
$$ \int_{S^1} ev^*(\mu _2). \eqno(4.10)$$
(One may think of the $S^1$
variable as introducing a periodic
time dependence and hence considering  time-dependant gauge
potentials and
gauge transformations). Of course on  $S^1(\cal A)$
the bundle 1-gerbe is trivial.
So the corresponding 2-cocycle is cohomologically
trivial. But the anomaly still manifests itself since the
2-form $\omega _2$ of equation (4.8) on
$S^1(\cal A) $ is not invariant under the action of $S^1( \cal G)$.
The corresponding anomaly can be obtained from (4.9).

Another interesting bundle 1-gerbe on
 ${\cal A}/{\cal G}$ is given by the following construction. Let
$\gamma $ be a closed
path in $M$ starting and ending at $x_0$. then the holonomy along $\gamma $
gives a
map:
$$
h: {\cal A}/{\cal G} \longrightarrow G.
\eqno(4.10)
$$
Therefore we have $h^*(\omega_3) \in H^3({\cal A}/{\cal G}, Z)$
where $\omega_3$ is the form (2.10).
Notice that (4.10) also gives the obvious map
$$
\tilde h: S^1(\cal A/\cal G) \longrightarrow S^1(G)
\eqno(4.11)
$$
As explained in section 2, the extension of the loop group $S^1(G)$
defined by $\omega_3$
is a  bundle 1-gerbe $Q_2\rightarrow S^1(G)$
on the compact Lie group $G$.
 Pulling back  $Q_2$ on $S^1(G)$  to
$S^1(\cal A/\cal G) $ using $\tilde h$, equation
(4.11), we obtain a
bundle 1-gerbe on
${\cal A}/{\cal G}$. We believe that this bundle
$1-gerbe$ is stably isomorphic
to $Q_1$ (see \cite{cmm} for the notion of stable isomorphism)
but do not have a proof as yet.

\section{ 3-Cocycles and bundle 2-gerbes}

In \cite{Jac} Jackiw argued that a non-vanishing 3-cocycle in
quantum field theories
 is a measure of non-associativity. There is a simple
quantum mechanical example,
arising from a point particle with charge $e$, at a point $\overrightarrow r$
moving in an external magnetic field
$\overrightarrow B$, which is not necessarily divergence free.
Geometrically speaking the Bianchi
identity fails. Non-associativity is however a paradoxical
interpretation in an operator algebra and the correct mathematical
point of view is to realise that when the Jacobi identity breaks
down in the sense that  the Lie triple brackets give a 3-cocycle,
one is dealing with an obstruction
to the existence of a Lie algebra extension.
Motivated by these quantum field theory examples,
in \cite{CarGruHurLan} and \cite{Car}, 3-cocycles as
obstructions to the existence of an extension of one Lie
algebra by another were
derived by a $C^*$-algebra method. It was found that the underlying
reason for the occurence of this cocycle in chiral  gauge theories
is that the equal time formalism is
 too singular in $3+1$ dimensions to permit
the definition of a consistent Lie algebra of canonical fields.
These anomalies do not as yet have a geometric interpretation but
it is tempting to speculate that
the notion of a bundle 2-gerbe may provide such an interpretation.
To see how this might work we consider an example for a  four
dimensional manifold $M$.

Henceforth, we suppose $M=S^4$, that $P$ is a principal
bundle over $M$, $\cal A$ is the irreducible connection space,
 $\cal G$ the gauge
group. We constructed in subsection  4.1 above
a
degree 4 cohomology class $[\Theta _4]$ in $H^4({\cal A}/{\cal G})$.
The results
 of section 3 show that $\Theta_4$ defines the   tautological
 bundle 2-gerbe
 on ${\cal A}/{\cal G}$.
On the other hand
the transgression of $\Theta_4$, say
 $\mu_3$ is a 3-cocycle on the Lie algebra of the gauge
group. As such
it may be thought of as an obstruction to the
existence of a Lie algebra extension.

However using the tautological construction
we can think of $\Theta_4$ as giving
rise to  an extension  of the loop
group of $\cal G$,
$S^1 ({\cal G})$.
To see this  we fix a base
point in each connected
component of ${\cal A}/{\cal G}$ and $\cal G$,.
Next  we consider the transgression:
$$
H^3({\cal G}, \mbox{Map}({\cal A}, R))\longrightarrow H^2(S^1 ({\cal G}),
\mbox { Map}
({\cal A}, R)).
\eqno(5.1)
$$
Under this map the cocycle $\mu_3 \in H^3({\cal G},
\mbox{Map}({\cal A}, R))$ gives a 2-cocycle $\Lambda$ on $S^1 ({\cal G}$
with coefficients in
 Map$({\cal A}, R)$.
Now $\Lambda$
defines an extension of $S^1 (L({\cal G}))$ by the abelian Lie algebra
$Map({\cal A}, R)$.

\noindent{\bf Proposition 5.1}:
 The 3-cocycle $\mu_3$ gives rise to an
extension of the Lie algebra
of the loop group of the gauge group, $S^1 ( L({\cal G }))$. The new Lie
bracket on $S^1 (L({\cal G}))$
is given by the following formula:
$$[(\epsilon_1, f_1), (\epsilon_2, f_2)]= ([\epsilon_1, \epsilon_2],
\epsilon_1.f_2-\epsilon_2.f_1+\Lambda(\epsilon_1,\epsilon_2))
\eqno(5.3)
$$
 where $[\epsilon_1,\epsilon_2]$ is the pointwise Lie bracket
and $\epsilon_i.f_j$ is the induced action of the Lie
algebra of the gauge group on functions on $\cal A$.

It would be interesting to understand whether the
3-cocycle studied in \cite{CarGruHurLan}
can be given a geometric understanding using the methods of this
paper.

\medskip
\noindent{\bf Remark}.
Interesting examples of de Rham forms on ${\cal A}/{\cal G}$
arise from the study of the descent equations \cite{Zum}
which are an example of a general approach
to constructing secondary characteristic classes due to Chern and Simons.
The Chern-Simons classes on $\cal A$
studied in \cite{Zum} are closed and $\cal G$ invariant and
so push down to forms on ${\cal A}/{\cal G}$.
They are not, however,  pull-backs of closed forms on this quotient space.
In the case of a 3-dimensional
base manifold $M$, the appropriate Chern-Simons 3-form on $\cal A$
may be written as $d\rho$ where $\rho$
 defines, using the transgression procedure of
Section 4, a two cocycle on the Lie algebra of the gauge group
 cohomologous to the
Faddeev-Mickelsson two cocycle.
This raises the question of whether the Chern-Simons forms of \cite{Zum}
pushed down to forms on ${\cal A}/{\cal G}$, are related in any way to the
class $[\Theta_3]$ of the bundle 1-gerbe of Section.
We have not been able to find such a relationship.

\end{small}
\end{document}